**Structural and Nanochemical Properties of AlO$_x$ Layers in Al/AlO$_x$/Al-Layer Systems for Josephson Junctions**

S. Fritz[1*], L. Radtke[2], R. Schneider[1], M. Luysberg[3], M. Weides[2,4], and D. Gerthsen[1*]

[1]Laboratory for Electron Microscopy, Karlsruhe Institute of Technology, 76131 Karlsruhe, Germany

[2]Physikalisches Institut, Karlsruhe Institute of Technology, 76131 Karlsruhe, Germany

[3]Ernst Ruska Center for Microscopy and Spectroscopy with Electrons and Peter Grünberg Institute 5, Forschungszentrum Jülich, 52425 Jülich, Germany

[4]James Watt School of Engineering, University of Glasgow, G12 8LT, Glasgow, United Kingdom

[*]Correspondence should be addressed to D.G. (email: dagmar.gerthsen@kit.edu) or S.F. (email: stefan.fritz@kit.edu)

**ABSTRACT**

The structural and nanochemical properties of thin $AlO_x$ layers are decisive for the performance of advanced electronic devices. For example, they are frequently used as tunnel barriers in Josephson junction-based superconducting devices. However, systematic studies of the influence of oxidation parameters on structural and nanochemical properties are rare up to now, as most studies focus on the electrical properties of $AlO_x$ layers. This study aims to close this gap by applying transmission electron microscopy in combination with electron energy loss spectroscopy to analyze the structural and nanochemical properties of differently fabricated $AlO_x$ layers and correlate them with fabrication parameters. With respect to the application of $AlO_x$ as tunnel barrier in superconducting Josephson junctions, $Al/AlO_x/Al$-layer systems were deposited on Si substrates. We will show that the oxygen content and structure of amorphous $AlO_x$ layers is strongly dependent on the fabrication process and oxidation parameters. Dynamic and static oxidation of Al yields oxygen-deficient amorphous $AlO_x$ layers, where the oxygen content ranges from $x = 0.5$ to $x = 1.3$ depending on oxygen pressure and substrate temperature. Thicker layers of stoichiometric crystalline $\gamma$-$Al_2O_3$ layers were grown by electron-beam evaporation of $Al_2O_3$ and reactive sputter deposition.

# I. INTRODUCTION

Aluminum oxide ($AlO_x$) layers are important components of several state-of-the-art electronic devices and decisive for their electronic properties. Thin $AlO_x$-tunnel barriers with a thickness of ~2 nm are widely used in $Al/AlO_x/Al$-based Josephson junctions (JJs) for superconducting electronic devices like superconducting quantum bits, single-electron transistors, single-photon detectors, radiation detectors and superconducting quantum interference devices in magnetometers [1–7]. Amorphous $AlO_x$ layers with a thickness of a few nanometers are used as gate dielectrics in high-gain graphene field-effect transistors [8, 9], as gate oxide in III-V compound semiconductor-based field-effect transistors [10, 11] or as layers in non-volatile resitive switching random access memories [12, 13]. The structural and nanochemical properties of $AlO_x$ layers have a significant influence on the performance of these devices. For example, $AlO_x$-layer thickness variations and structural defects in $AlO_x$-tunnel barriers of JJs cause noise and limit the detection sensitivity of superconducting interference devices and coherence times in quantum bits [14–20].

Amorphous $AlO_x$ layers are fabricated by static or dynamic oxidation of electron-beam evaporated Al layers in deposition systems with a base pressure in the high-vacuum (HV) regime. Static oxidation is performed with a fixed $O_2$-partial pressure whereas a constant $O_2$ flow is used under dynamic oxidation conditions. The oxidation process is a self-limiting process with fast oxidation during the initial stage, followed by saturation of oxide-layer growth and decreasing oxidation rate towards zero which can be described by the Cabrera-Mott model [21–24]. Self-limiting oxide growth is advantageous for $AlO_x$-tunnel barriers in JJs because it provides a high degree of reproducibility, which is mandatory for large-scale fabrication. Numerous studies were performed to correlate oxidation conditions and critical current in JJs. Specifically, the influence of temperature [25] and oxygen pressure during static oxidation [26–28] and both combined [29–31] were studied. Variations of the critical current are usually attributed to the variation of the $AlO_x$-layer thickness [32, 33]. However, $AlO_x$ composition variations and changes of the Al-O coordination, i.e. the average number of bonds between Al and O, could also affect the resistivity of the tunnel barrier and require atomic-scale analyses of the $AlO_x$ composition. Only few studies of nanochemical $AlO_x$ properties exist. For example, using nano-beam electron diffraction in a transmission electron microscope Zhang *et al.* [34] demonstrated oxygen deficiency at $Al/AlO_x$ interfaces of $AlO_x$ tunnel barriers.

To obtain information on the properties of the amorphous $AlO_x$ layers on the nanoscale, this work is concerned with the correlation of the $AlO_x$-oxidation conditions with the structural and nanochemical $AlO_x$ properties investigated by transmission electron microscopy (TEM) combined with electron energy loss spectroscopy (EELS). We have investigated in particular the oxygen concentration and potential conditions that can lead to crystalline $Al_2O_3$. The latter goal is motivated by the observation that quantum bits containing an epitaxially grown crystalline $Al_2O_3$ tunnel barrier show a reduced density of two-level systems [35] and reduced coupling strength [36]. These observations emphasize once more the importance of the structural and nanochemical properties of $AlO_x$-tunnel barriers with respect to the optimization of superconducting devices based on JJs.

Our previous work on the optimization of $Al/AlO_x/Al$-layer systems for JJs has demonstrated that $AlO_x$-tunnel barriers with highly homogeneous thickness can be obtained on an epitaxial Al(111) lower electrode layer grown on Si(111) substrates [37]. Epitaxial Al(111) is essential to provide a flat surface which is well suited for oxidation and leads to significantly reduced thickness variations of the amorphous $AlO_x$-tunnel barrier. In the present work, we focus on the oxidation parameters (substrate temperature and oxygen pressure), which are correlated with structural and nanochemical $AlO_x$ properties in different $Al/AlO_x/Al$-layer systems. Most of our growth experiments were performed in a standard HV electron-beam physical vapor deposition system. $Al/AlO_x/Al$-layer systems were also fabricated by reactive sputter deposition. Structural and nanochemical properties were investigated by high-resolution (HR)TEM in combination with EELS. We will demonstrate that the oxygen content and the structure of the AlO$_x$ layer is strongly dependent on the fabrication process and the applied oxidation parameters. Dynamic and static oxidation of Al in an oxygen environment yields oxygen-deficient amorphous AlO$_x$ layers, where the stoichiometry ranges from AlO$_{0.5}$ to AlO$_{1.3}$ depending on oxygen pressure and substrate temperature. EELS analyses demonstrate that the Al-O bonding characteristics change for different substrate temperatures indicating a structural change towards crystalline structures for oxidation temperatures above 200 °C.

**II. METHODS**

The majority of the investigated Al/AlO$_x$/Al-layer systems were fabricated in a single-chamber *MEB 550S* (*PLASSYS Bestek, Marolles-en-Hurepoix, FR*) electron-beam physical vapor deposition system with a base pressure of 10$^{-7}$ mbar in the HV regime. It is equipped with a sample-plate heater for heating the substrate up to 700 °C, an UV lamp for oxidation enhancement and a *Kaufman* source to generate an Ar/O-plasma for removing carbonaceous contamination from the substrate or plasma-enhanced oxidation.

The fabrication process is in detail described in the Supplementary Information and in [37, 38]. Briefly, the 100 nm thick lower Al layer was deposited on cleaned Si(001) and Si(111) substrates by electron-beam evaporation at a chamber pressure of 8 – 12·10$^{-8}$ mbar. Epitaxial Al growth was obtained on Si(111) substrates pretreated by an HF-dip and 700 °C annealing in combination with Al-deposition temperatures ≤ 100 °C and Al-deposition rates ≥ 0.5 nm/s, which results in Al-layer thickness variations of less than ± 2 nm over lateral distances of ~15 μm [37]. After Al deposition, different oxidation methods and parameters were applied to form the AlO$_x$ layer. By varying the oxidation parameters, we aim to correlate the oxidation parameters with structural and nanochemical properties of different AlO$_x$ layers. We note that many studies consider the oxygen exposure, i.e. the product of oxidation time and oxygen pressure, as a decisive parameter [28, 32]. However, we consider the effect of oxidation time to be negligible as long as the oxidation time is long enough to reach the saturation regime for the $AlO_x$ thickness [25, 26, 29]. The timeframe in which the saturation regime and the self-limiting thickness of the $AlO_x$-layer is reached can vary considerably depending on the experimental setup. For example, the self-limiting thickness can be reached within less than 4 min as demonstrated by Jeurgens et al. [25], but it can also take more than 200 min [26]. To estimate the timeframe for our experimental setup, we fabricated three samples with identical oxidation temperatures of 250 °C and oxygen pressures of 0.3 mbar but with varying oxidation times of 20 s, 5 min and 30 min. The sample with an oxidation time of 20 s shows an $AlO_x$ thickness of 1.14±0.10 nm. It increases to 1.59±0.11 nm for an oxidation time of 5 min and essentially remains constant for longer oxidation times (30 min, $AlO_x$-thickness of 1.57±0.12 nm). Thus, we conclude that the saturation regime is reached within less than 5 min and oxidation times exceeding 5 min will not have an impact on the structural and nanochemical properties of the $AlO_x$ layers.

Table I summarizes the conditions for differently fabricated samples. Dynamic oxidation for 12.5 min at room temperature was applied for samples EBPlas and EBPlas-UV grown on Si(001) substrates with a constant oxygen flow of 10 and 12.7 sccm. For EBPlas-UV, additional UV-light illumination was used to enhance the oxidation process. Static oxidation was applied for samples deposited on Si(111), where the oxygen pressure was varied between 0.3 mbar and 9.5 mbar. Substrate temperatures during oxidation were either 70 °C or 250 °C (cf. Table I with sample denotations EBPlas-70|0.3, EBPlas-70|9.5, EBPlas-250|0.3 and EBPlas-250|9.5). Independent of oxidation type and parameters, the oxidation process is self-limiting and yields $AlO_x$ layers with a thickness below 2 nm [32, 37].

| sample | oxidation temperature [°C] | O$_2$ pressure [mbar] | oxidation time [min] |
|---|---|---|---|
| Si (001) substrate and dynamic oxidation | | | |
| EBPlas | 30±5 | 0.015±0.01  O$_2$ flow: 10 sccm | 12.5 |
| EBPlas-UV | 30±5 | 0.02±0.01  O$_2$ flow: 12.7 sccm | 12.5 |
| HF-cleaned Si (111) substrate and static oxidation | | | |
| EBPlas-70\|0.3 | 70±10 | 0.3±0.1 | 12.5 |
| EBPlas-70\|9.5 | 70±20 | 9.5±0.5 | 5 |
| EBPlas-250\|0.3 | 250±20 | 0.3±0.1 | 5 |
| EBPlas-250\|9.5 | 250±50 | 9.5±0.5 | 6 |

**Table I**. Oxidation conditions for the samples fabricated in the *Plassys MEB*.

However, thicker AlO$_x$ layers are required for quantitative chemical analysis due to limitations of the used transmission electron microscope with an electron-beam diameter of 1.8 nm (cf. Supplementary Information). For this reason, the samples listed in Table I were produced with 15 – 20 nm AlO$_x$ layers by iterative oxidation. After the first oxidation step, 1 nm Al is deposited and oxidized under the same conditions. This process is repeated for up to 15 times using the same oxidation conditions for each iteration until the desired AlO$_x$-layer thickness is reached. In the last step, a 100 nm thick upper Al layer was deposited using the same Al-deposition parameters as for the lower Al layer.

An additional sample, denoted as EBPlas-250|0.3|JJ, was fabricated with identical oxidation parameters as EBPlas-250|0.3 (250 °C and 0.3 mbar) but with only a single oxidation step to obtain a ~2 nm thin AlO$_x$-tunnel barrier to analyze differences between ~20 nm thick AlO$_x$ layers and thin AlO$_x$-tunnel barriers.

For comparison, two samples were deposited in other systems. The sample EBLes was fabricated in a different electron-beam physical vapor deposition system (*PVD 75, Kurt J. Lesker Company, Hastings, UK*), which allows direct AlO$_x$ deposition by evaporation of Al$_2$O$_3$ pellets. Sample ReSput was grown in a home-built sputter deposition system. An Ar plasma was used for sputter deposition of the lower and upper Al layer, while an Ar/O-plasma (9:1 mixture) was used for sputter deposition of the $AlO_x$ layer.

Cross-section specimens for TEM were prepared for all samples by conventional mechanical preparation techniques as described by Strecker *et al.* [39] using $Ar^+$-ion milling with a *Gatan 691 PIPS* (*Gatan Inc., Pleasanton, USA*) as final preparation step.

Transmission electron microscopy (TEM) was performed with an *FEI Titan³ 80-300* (*Thermo Fisher Scientific, Waltham, USA*) operated at 300 kV. The microscope is equipped with a Gatan imaging filter *Tridiem HR 865* (*Gatan Inc., Pleasanton, USA*) and an aberration corrector in the imaging lens systems. Structural analyses of crystalline regions in the $Al/AlO_x/Al$ layers were performed by comparing two-dimensional (2D) Fourier-transform patterns of HRTEM images with simulated diffraction patterns using the *JEMS* software [40].

EELS was performed in the scanning (S)TEM mode using a self-written acquisition script including binned-gain averaging [41] to enhance the signal-to-noise ratio. Chemical composition quantification is based on the k-factor method [38, 42, 43] using crystalline $\gamma$- and $\alpha$-$Al_2O_3$ as reference materials.

EELS measurements were also performed with a probe-corrected *FEI Titan 80-300* (*Thermo Fisher Scientific, Waltham, USA*) equipped with a Gatan imaging filter *Tridiem 866ERS* (*Gatan Inc., Pleasanton, USA*) at the Ernst Ruska Center for Microscopy and Spectroscopy with Electrons (Forschungszentrum Jülich, Jülich, GER) using the StripeSTEM technique [44]. EELS acquisition conditions and the subsequent analysis process are in detail described in ref. [38] and in the Supplementary Information.

### III. RESULTS AND DISCUSSION

**A. Morphology and structure of the $AlO_x$ layers**

The morphology of the AlO$_x$ layers is predominantly determined by the morphology of the lower Al layer. Our previous work has demonstrated that thickness variations of AlO$_x$-tunnel barriers are directly correlated with thickness variations of the lower Al layer, which can be minimized by epitaxial growth of Al(111) on Si(111). Under optimized conditions [37] thickness fluctuations of the AlO$_x$-tunnel barrier are reduced to ± 0.11 nm over lateral distances of about 15 µm. Figure 1 shows a cross-section HRTEM image of the tunnel-barrier region of EBPlas-250|0.3|JJ, where small AlO$_x$-thickness fluctuations are mainly caused by atomic steps at the Al(111)/AlO$_x$ interface. We point out that the crystalline orientation of the lower Al layer is transferred to the upper Al layer despite the presence of the amorphous AlO$_x$-tunnel barrier. Due to a slight tilt of the upper Al layer by about 0.5° compared to the lower Al layer, the step density is not identical at the upper and lower interface and atomic steps do not occur at the same lateral position. This inevitably results in thickness variations of the AlO$_x$ layer, which are, however, small compared to thickness variations caused by grain-boundary grooving in polycrystalline Al-electrode layers. The sharp Al/AlO$_x$ transition seems to occur within one Al(111) lattice plane (0.23 nm).

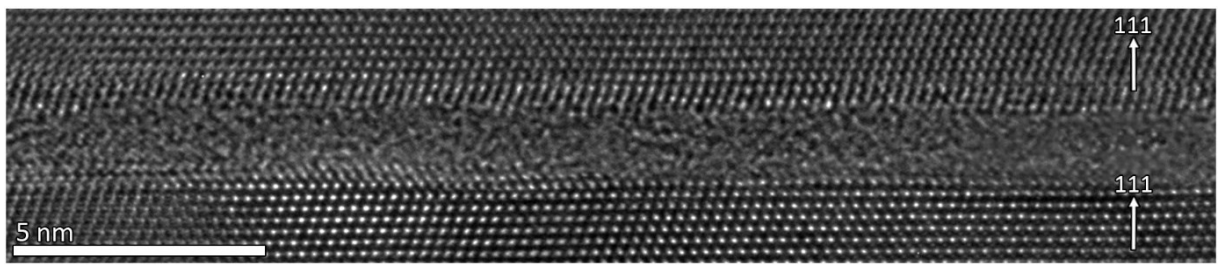

**Figure 1**. HRTEM image of EBPlas-250|0.3|JJ. The AlO$_x$ layer was obtained by static oxidation at 250 °C with an oxygen pressure of 0.3 mbar. The lower Al(111) layer was epitaxially grown on Si(111) at 100 °C with an Al-deposition rate of 0.5 nm/s.

For quantification of the chemical composition, thicker AlO$_x$ layers are required because the minimum electron-beam diameter of ~1.8 nm in our transmission electron microscope does not allow reliable composition analyses in thin tunnel barriers (cf. Supplementary Information). Figure 2 shows HRTEM images EBPlas (Figure 2a) and EBPlas-250|9.5 (Figure 2d) with

thicker AlO$_x$ layers. Iterative Al deposition/oxidation was applied yielding AlO$_x$ thicknesses between 15 and 25 nm.

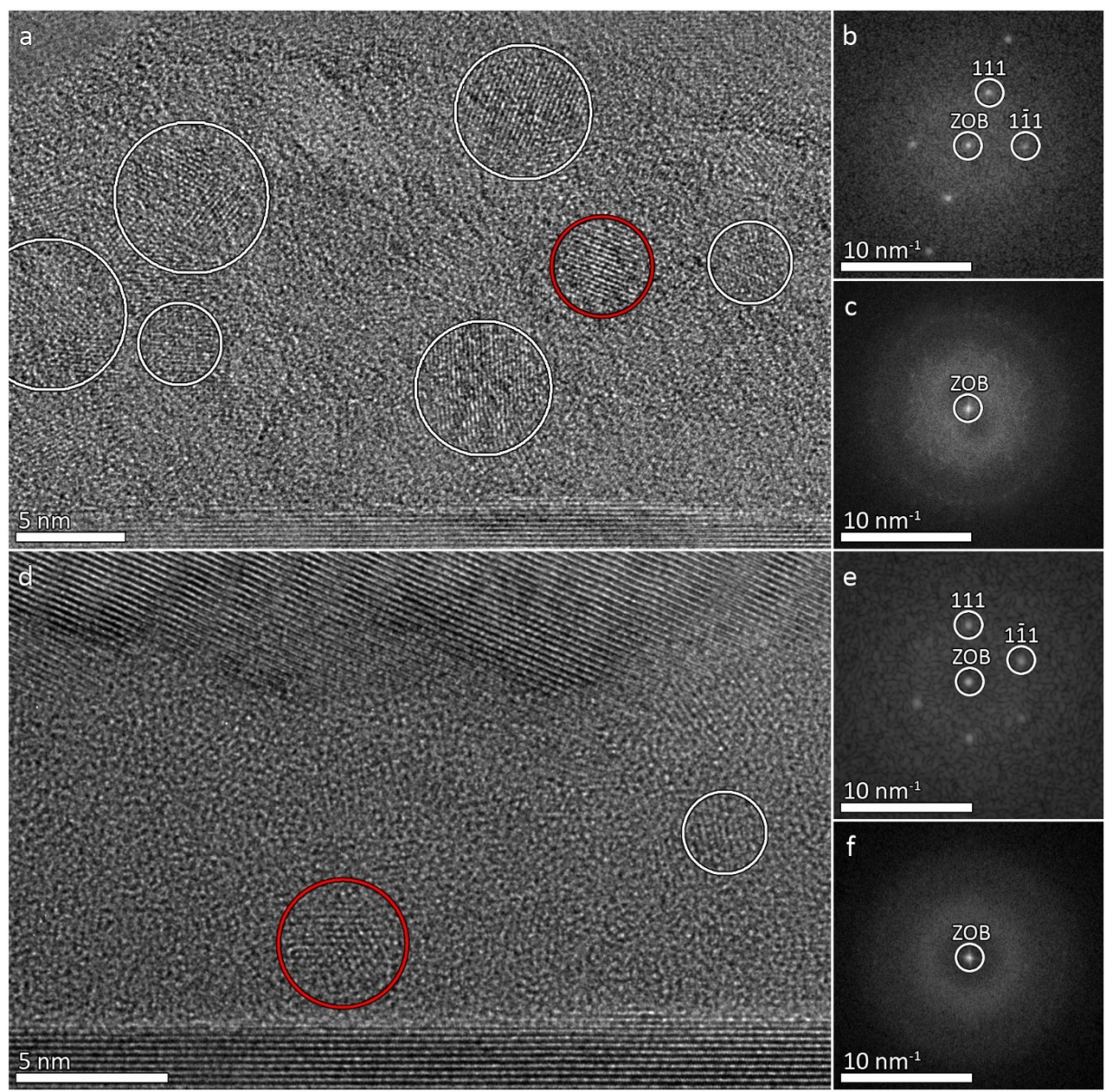

**Figure 2**. HRTEM images of the AlO$_x$ layer in (a) EBPlas and (d) EBPlas-250|9.5 with corresponding FT patterns of (b,e) nanocrystalline inclusions marked by red circles and (c,f) amorphous regions. Further crystalline Al inclusions are marked by white circles. The zero-order beam (ZOB) and two Bragg reflections are indexed in the FT patterns of crystalline inclusions in (b) and (e).

The AlO$_x$ layer of EBPlas contains nanocrystalline inclusions with sizes between 2 and 4 nm (circles in Figure 2a) embedded in amorphous AlO$_x$. Calculated two-dimensional (2D) Fourier-transform (FT) patterns of the inclusions (cf. Figure 2b) perfectly agree with simulated diffraction patterns of pure Al (face-centered cubic structure, space group Fm-3m, lattice parameter a = 4.06 Å). The FT pattern in Figure 2c was obtained from an amorphous region and does not show any Bragg reflections as expected. The crystalline inclusions are distributed randomly throughout the whole AlO$_x$ layer in EBPlas. Their 'area density' was determined to be 19 % by measuring the ratio between amorphous and crystalline regions of the AlO$_x$ layers over a lateral distance of about 1 µm using several HRTEM images.

Figure 2d shows an HRTEM image of EBPlas-250|9.5 with corresponding FT patterns of a crystalline Al inclusion (Figure 2e) and an amorphous region (Figure 2f). The AlO$_x$ layer also contains crystalline Al inclusions, but with a reduced fraction of only 3 %. HRTEM images of the other four samples produced by static and dynamic oxidation (not shown here) show similar features. Only the content of crystalline Al inclusions differs and is listed in Table II together with the oxygen pressure during static/dynamic oxidation.

Crystalline Al inclusions are only observed in multiple oxidized AlO$_x$ layers, where an iterative Al-deposition/oxidation process (cf. Section II) was used. For each iteration, an Al layer with an intended thickness of 1 nm was deposited with a nominal deposition rate of 0.1 nm/s. This results in a very short deposition time and the total amount of deposited Al can vary due to fluctuation of the deposition rate or delays during the closure of the mechanical shutter. We assume that the Al does not grow as a homogeneous 1 nm thick layer, but rather forms islands with different height, which locally may prevent complete oxidation of the islands.

| sample | fraction of crystalline inclusions | O₂ pressure [mbar] |
|---|---|---|
| **EBPlas** | 19% | 0.015 |
| **EBPlas-UV** | 14% | 0.02 |
| **EBPlas-70\|0.3** | 8% | 0.3 |
| **EBPlas-70\|9.5** | 3% | 9.5 |
| **EBPlas-250\|0.3** | 8% | 0.3 |
| **EBPlas-250\|9.5** | 3% | 9.5 |

**Table II**. Fraction of crystalline Al inclusions embedded in amorphous $AlO_x$ layers and corresponding $O_2$ pressure during dynamic/static oxidation in the *Plassys MEB*.

The data in Table II clearly show a correlation between oxygen pressure during the oxidation and fraction of crystalline Al inclusions. Dynamic oxidation of EBPlas with 10 sccm, which corresponds to an oxygen pressure of 15 µbar, leads to the highest fraction of Al inclusions (19 %). With increasing oxygen pressure to 20 µbar (EBPlas-UV), the fraction of Al inclusions is reduced to 14 %. It decreases to only 3 % for the highest pressure of 9.5 mbar. Variation of the oxidation temperature does not have a measurable effect on the fraction of Al inclusions.

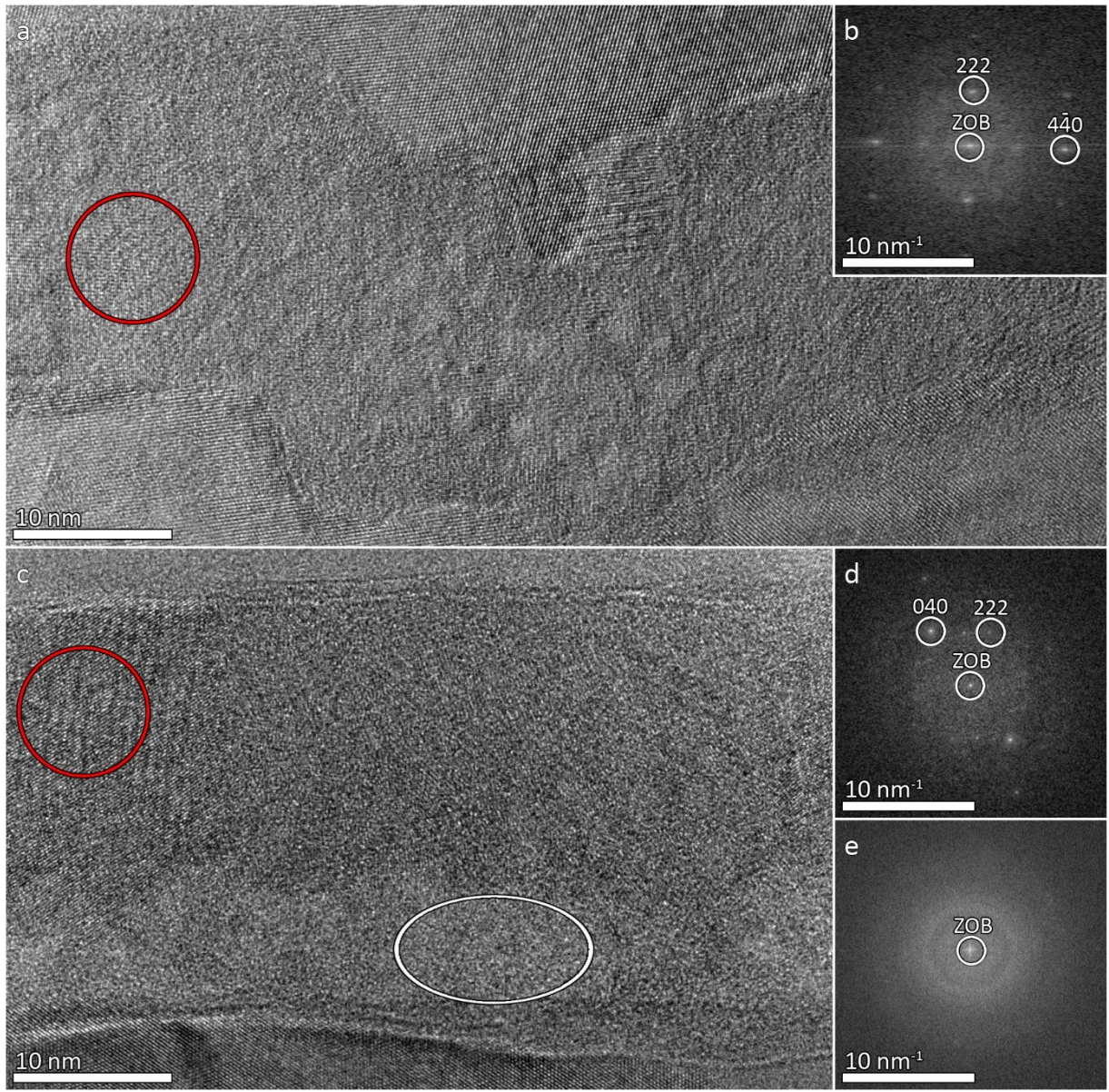

**Figure 3**. HRTEM images of the AlO$_x$ layer in (a) EBLes fabricated by electron-beam evaporation of Al$_2$O$_3$ and (c) ReSput fabricated by reactive sputter deposition with an Ar/O plasma with corresponding FT pattern of (b,d) crystalline (red-encircled) and (e) amorphous (white-encircled) regions. The zero-order beam (ZOB) and two indexed Bragg reflections are marked in the FT pattern.

HRTEM images of the AlO$_x$ layers fabricated in the other two deposition systems are shown in Figure 3. Their morphology differs from the samples shown in Figure 2. The pronounced corrugation of the lower Al/AlO$_x$ interface in EBLes (Figure 3a) is associated with nanoscaled Al grains with an average size of only 54 ± 23 nm in the lower Al layer. We attribute the small Al-grain sizes to the comparatively high base pressure of the deposition system (~10$^{-6}$ mbar), because residual gases like oxygen can have strong impact on the Al-grain size and Al-growth process [45, 46]. The AlO$_x$ layer of EBLes was deposited by electron-beam evaporation of Al$_2$O$_3$ pellets. Despite the poor morphology, i.e. strong corrugation of the lower interface, a crystalline Al-oxide layer was obtained. Structural analysis (Figure 3b) yields the cubic defect spinel structure (space group Fd-3m, lattice parameter a = 7.91 Å [47]) that corresponds to the γ-Al$_2$O$_3$ phase. The γ-Al$_2$O$_3$ layer consists of small crystalline grains with sizes of 10 – 50 nm and different crystallographic orientations.

The AlO$_x$ layer of ReSput (cf. Figure 3c) was fabricated by reactive sputter deposition using an Ar/O-plasma (9:1 mixture) and a pure Al target. The corrugation of the lower Al/AlO$_x$ interface is attributed to the sputter process or Ar-plasma cleaning between Al deposition and oxidation. Remarkably, the AlO$_x$ layer is subdivided into an amorphous lower sublayer with an average thickness of 8.6 ± 2.1 nm, which exhibits a slightly brighter contrast in Figure 3b (see also FT pattern in Figure 3e from the white-encircled region). The upper sublayer is crystalline and consists of the cubic defect spinel structure of γ-Al$_2$O$_3$ (FT pattern in Figure 3d of the red-encircled region in Figure 3c). Such a transition from amorphous AlO$_x$ to polycrystalline γ-Al$_2$O$_3$ within the oxide layer was only observed in ReSput and could be induced by two different mechanisms. From a thermodynamic point of view, the oxide layer strives to minimize the total Gibbs free energy of the system. According to theoretical considerations by Jeurgens *et al.* [48] and experiments by Reichel *et al.* [49] there is a critical thickness upon which the crystalline phase is thermodynamically preferred and thus a transition from the amorphous to the crystalline phase takes place. The critical thickness for the transition depends on the crystallographic orientation of the Al surface and temperature. Jeurgens *et al.* calculated values of 0.3 nm, 2 nm and 4 nm for Al(111), Al(100) and Al(110) surfaces at room temperature. The critical thickness for higher-index Al surfaces, as present in ReSput, could be even larger but still in the same order of magnitude and compatible with the measured thickness of ~ 9 nm. A second mechanism could be responsible for the amorphous-to-crystalline transition. After sputter deposition of the lower Al layer with a pure Ar plasma, additional oxygen was added to the plasma. As usual for sputter deposition, a pre-sputter process with closed shutter was performed to homogenize the deposition rate and remove possible contamination from the sputter target prior to the actual deposition process. Although the material cannot be deposited on the sample, oxygen atoms from the plasma still can reach the Al surface and oxidize it. The oxidation condition is then comparable to plasma-enhanced static oxidation, which was also applied in our *Plassys* system and resulted in a ~5 nm thick amorphous $AlO_x$ layer [37]. Hence, the measured thickness of ~9 nm of the amorphous $AlO_x$ sublayer in ReSput excludes that it is solely formed by the second process. We suggest that the first few nm of the amorphous sublayer is formed during pre-sputtering by oxidation from the Ar/O plasma. After opening the shutter, the $AlO_x$ layer continues to grow in an amorphous structure until the thermodynamic driving force is high enough to cause the transition to the crystalline γ-phase. The layer then continues to grow in the crystalline phase until the reactive sputter process is stopped.

A transition from the amorphous to the crystalline phase was only observed in sample ReSput fabricated by reactive sputter deposition, although such a transition was previously reported to occur during static oxidation in an ultrahigh vacuum (UHV) system [49]. We did not observe an amorphous/crystalline transition in our static-oxidized $AlO_x$ layers and assume that it is prevented by the iterative Al-deposition/oxidation process and the lack of a UHV environment in our deposition system. Residual adsorbates on the Al surface may disturb the crystallization process and thus prevent the transition to the crystalline phase. However, our sputter deposition system is also not operated under UHV conditions, but oxidation conditions differ substantially from static and dynamic oxidation. The Ar/O plasma contains energetically activated O ions instead of $O_2$ molecules. O ions react more strongly with Al atoms resulting in a higher oxygen content, which then is expected to lower the energy barrier for the transition to stoichiometric crystalline $Al_2O_3$.

**B. Chemical composition, Al-O bonding characteristics and near-range order of $AlO_x$**

The present study is motivated by the goal to achieve more detailed information on the correlation between deposition conditions, chemical composition and structural characteristics of differently fabricated AlO$_x$ layers, which can be obtained from EELS and energy-loss near-edge structure (ELNES) analyses of the Al-L$_{2,3}$ and O-K ionization edges acquired in a transmission electron microscope (for experimental details see Supplementary Information).

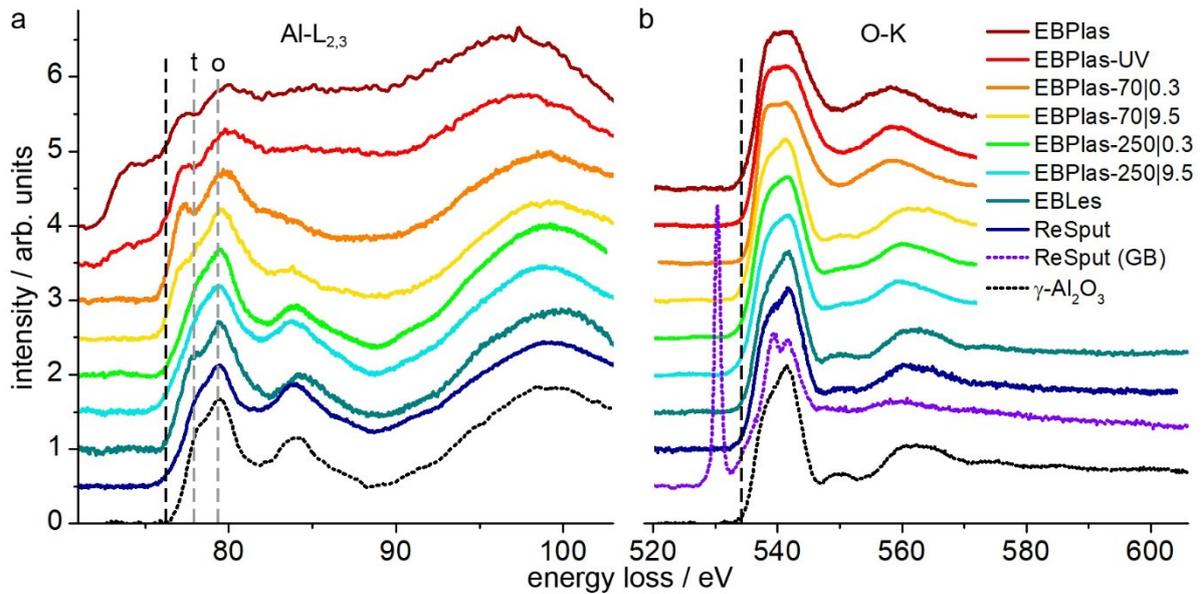

**Figure 4**. EELS spectra of the (a) Al-L$_{2,3}$ and (b) O-K ionization edges obtained from AlO$_x$ layers fabricated in different systems with different oxidation parameters (see legend). A γ-Al$_2$O$_3$ reference spectrum is also included (black dotted line). The spectrum denoted as ReSput (GB) was acquired from a grain-boundary region between two γ-Al$_2$O$_3$ grains in ReSput. Al-L$_{2,3}$ and O-K edge onsets are marked by black dashed lines. Gray dashed lines labeled 't' and 'o' mark peaks corresponding to tetrahedral- and octahedral-coordinated Al sites of γ-Al$_2$O$_3$.

Figure 4 shows EELS spectra of the AlO$_x$ layers of all samples with an additional reference spectrum of pure γ-Al$_2$O$_3$ (*Carl Roth GmbH + Co KG, Karlsruhe, GER*). All EELS spectra were acquired in amorphous regions of the AlO$_x$ layer to minimize the influence of the crystalline inclusions consisting of pure Al (cf. Figure 2). The spectrum of the Al-L$_{2,3}$ ionization edge (Figure 4a) of γ-Al$_2$O$_3$ (black dotted line) shows an edge onset at 76 eV followed by two peaks at 77.9 eV and 79.4 eV, which can be associated with tetrahedral- (4-fold) and octahedral- (6-fold) coordinated Al atoms (gray dashed lines), where Al atoms are surrounded by either 4 or 6 O atoms [50, 51]. The energy loss range from 81 eV to 89 eV contains characteristic features of intermediate-range order interactions [52]. The peak at 84 eV indicates a high degree of structural order as expected for crystalline material.

The EELS spectra of the differently fabricated $AlO_x$ layers show pronounced ELNES variations. The onset energy is reduced to 72.5 eV for EBPlas (brown line) and EBPlas-UV (red line), which indicates the presence of metallic Al [51]. This can be attributed to the high content of crystalline Al inclusions observed in HRTEM images, which makes it practically impossible to acquire spectra without the influence of the crystalline Al inclusions. The reduced content of Al inclusions in EBPlas-UV leads to a less intense 'pure Al' feature in the spectrum. All other spectra do not show any hint of pure crystalline Al, although HRTEM images of EBPlas-70|0.3 and EBPlas-250|0.3 also show a non-negligible amount of Al inclusions (8 %). The spectra also differ considerably in the intensity and position of the two peaks corresponding to tetrahedral- and octahedral-coordinated Al sites and the existence of the intermediate-range order peak at 84 eV. The lack of the latter is typically associated with an amorphous structure as observed for EBPlas, EBPlas-UV and EBPlas-70|0.3, where the oxidation took place at low temperatures (room temperature or 70 °C) and low oxygen pressures (0.3 mbar and below, cf. Table I). The tetrahedral peak shifts to lower and the octahedral peak to higher energies for these three samples. According to Bruley *et al.* [53], the ratio of Al atoms with tetrahedral and octahedral coordination can be calculated by the evaluation of peak intensities in crystalline $Al_2O_3$ phases. The evaluation of intensity of the two peaks in $\gamma\text{-}Al_2O_3$ yields indeed the expected values of 30 % tetrahedral- and 70 % octahedral-coordinated Al atoms. However, an amorphous phase also contains 1-, 2-, 3- and 5-fold coordinated atoms [34], preventing straightforward determination from EELS spectra. Moreover, the fraction of the differently coordinated Al atoms depends strongly on the density and stoichiometry of the $AlO_x$ layer [54]. Nevertheless, the pronounced tetrahedral peak indicates a shift towards low-coordinated atoms and lower oxygen content in EBPlas, EBPlas-UV and EBPlas-70|0.3 compared to the

crystalline γ-Al$_2$O$_3$ phase. With increasing oxygen pressure for EBPlas-70|9.5 (yellow line in Figure 4a), the tetrahedral peak decreases in intensity and a weak intermediate-range order peak appears. The spectra of the AlO$_x$ layers oxidized at high temperatures (EBPlas-250|0.3 and EBPlas-250|9.5, green and turquoise lines in Figure 4a) agree remarkably well (apart from a slightly broadened intermediate-range order peak) with the spectrum of crystalline γ-Al$_2$O$_3$, although HRTEM images do not indicate the presence of a long-range ordered (crystalline) structure. These observations demonstrate that the change of the ELNES features of differently fabricated amorphous AlO$_x$ materials reveals much more clearly changes on an atomic scale than HRTEM images. As expected, spectra of ReSput (spectrum acquired in the crystalline sublayer, blue line in Figure 4a) and EBLes (dark green line in Figure 4a) agree well with the γ-Al$_2$O$_3$ reference apart from a slightly broadened intermediate-range order peak. This is attributed to the nanocrystalline γ-Al$_2$O$_3$ structure of these samples and the resulting high concentration of grain boundaries, leading to structural disorder at grain boundaries.

The EELS spectra of the O-K ionization edge in Figure 4b reveal less pronounced ELNES features, but show the same trend as the Al-L$_{2,3}$ edge. The edge onset at 533 eV (black dashed line) is identical for all samples and is not affected by the presence of crystalline Al. The shape of the following intense main peak varies from a sharp peak with a maximum at 541 eV for the crystalline γ-Al$_2$O$_3$ to a flattened peak for the amorphous AlO$_x$ layers (EBPlas, EBPlas-UV and EBPlas-70|0.3). Two smaller broadened peaks at 550 eV and 563 eV occur in crystalline γ-Al$_2$O$_3$, whereas amorphous layers show only one peak at 558 eV. AlO$_x$ layers oxidized at high pressure and/or high temperature show a gradual transition between amorphous and crystalline features. The O-K edge of nanocrystalline γ-Al$_2$O$_3$ in EBLes and ReSput again is almost identical to γ-Al$_2$O$_3$.

Interestingly, the spectrum denoted as ReSput (GB) (purple dotted line in Figure 4b), which was acquired at a grain boundary between two γ-Al$_2$O$_3$ grains in ReSput, deviates drastically from the other spectra. A sharp and intense pre-peak at 530.6 eV arises and the main peak is split into maxima at 539.6 eV and 541.9 eV. This indicates a change of bonding between Al and O atoms. In fact, the ELNES can be associated to the presence of O-O bonds because the spectrum perfectly agrees with soft x-ray emission spectroscopy, x-ray absorption spectroscopy or EELS data of molecular oxygen [55–57]. The peak at 530.6 eV can be assigned to the excitation of the $\pi$ orbitals and the peaks at 539.6 eV and 541.9 eV to the excitation of $\sigma$ resonances [56]. We note that such ELNES features were only found at grain boundaries between $\gamma$-$Al_2O_3$ grains in EBLes and ReSput. The spectrum does not arise or change during electron-beam illumination and therefore does not result from electron-beam damage. This implies that the molecular oxygen is inherently present at the grain boundaries in $\gamma$-$Al_2O_3$ layers fabricated by reactive sputter deposition and electron-beam evaporation of $Al_2O_3$.

In addition to the analysis of ELNES fingerprints, the EELS spectra can be used to determine the chemical composition by the evaluation of the integrated intensities of the Al-$L_{2,3}$ and O-K edges (cf. Supplementary Information). Figure 5 shows the oxygen content x of the different $AlO_x$ layers with x = 1.5 corresponding to stoichiometric $Al_2O_3$. All amorphous layers are oxygen deficient as expected from previous experimental work [34, 38, 58] and simulations [54]. Dynamic oxidation with a low oxygen pressure of 15 µbar (EBPlas) shows an oxygen content of only x = 0.5. However, HRTEM images (cf. Figure 2a) of this sample show a considerable fraction of crystalline Al inclusions (19 %), which are randomly distributed within the $AlO_x$ layer and thus reduce the average oxygen content. Also, the Al-$L_{2,3}$ ELNES shows an edge onset at 72.5 eV indicating the presence of pure Al. We, therefore, conclude that the extremely low oxygen content is partially induced by the presence of crystalline Al. The oxygen content is increased to x = 1.1 by additional UV illumination during the oxidation. UV light enhances the dissociation rate of $O_2$ and generates energetically activated O ions with a reduced activation barrier for chemisorption [59]. The reduced content of crystalline Al inclusions (14 %) could also contribute to the increased oxygen content.

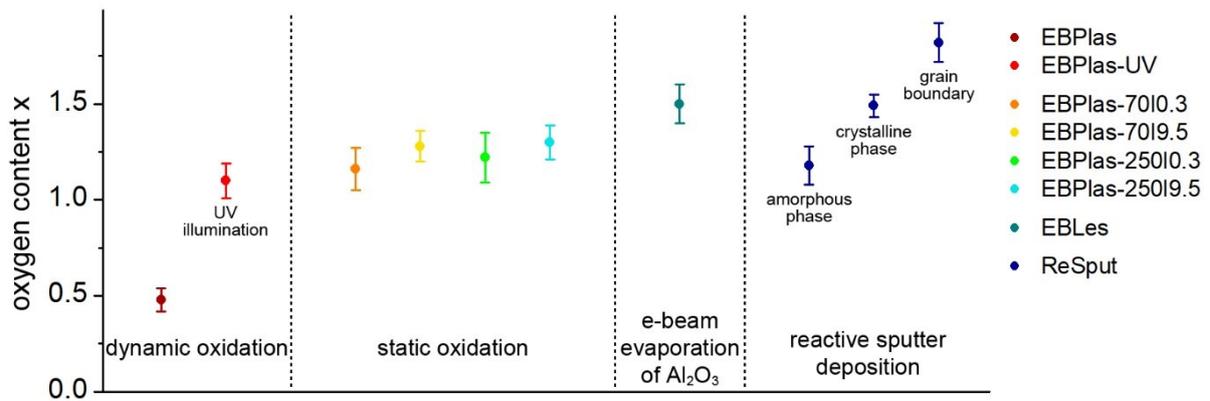

**Figure 5**. Oxygen content x of AlO$_x$ layers fabricated with different oxidation techniques and varying oxidation parameters.

Static oxidation further enhances the oxygen concentration depending on oxidation temperature and oxygen pressure. The fraction of crystalline inclusions in these samples is between 3 and 8 % (cf. Table II) and will only marginally affect the measured oxygen concentration. For a constant temperature of 70 °C, the oxygen content increases with pressure from x = 1.16 (0.3 mbar) to x = 1.28 (9.5 mbar). The same effect is observed at 250 °C. For a constant oxygen pressure, the oxygen content increases slightly with temperature. For example, at an oxygen pressure of 0.3 mbar the oxygen content increases from x = 1.16 (70 °C) to x = 1.22 (250 °C). Both effects can be used to maximize the O content which reaches x = 1.30 for EBPlas-250|9.5. EBLes and the crystalline sublayer of ReSput contain an oxygen content of x = 1.5 as expected for the stoichiometric γ-Al$_2$O$_3$ phase. An oxygen content of x = 1.18 was found for the amorphous sublayer of ReSput, which does not contain crystalline Al inclusions. Grain boundaries between two neighboring crystalline γ-Al$_2$O$_3$ grains in ReSput show an oxygen excess with x = 1.8. This high oxygen content is consistent with the existence of molecular O$_2$ at the GBs as demonstrated by the ELNES of the O-K ionization edge (cf. Figure 4b).

The most striking observation revealed by ELNES in differently fabricated $AlO_x$ layers is the gradual transition from an amorphous state to a state that shows crystalline features at high oxidation temperature and oxidation pressure, although HRTEM images still suggest purely amorphous $AlO_x$ (cf. Figure 2d for EBPlas-250|9.5) and all spectra were acquired in amorphous-looking regions without influence of crystalline Al inclusions. The transition is accompanied by an increasing O concentration which reaches a maximum of x = 1.3 in EBPlas-250|9.5. In this context, the work by Reichel *et al.* [49] can be again invoked who found the structure of $AlO_x$ on Al surfaces to be temperature dependent. Amorphous $AlO_x$ on Al(111) prevails at temperatures below 175 °C, whereas crystalline $\gamma$-$Al_2O_3$ layers were formed at higher oxidation temperatures. Thus, crystalline $AlO_x$ layers would be expected for our samples EBPlas-250|9.5 and EBPlas-250|0.3 on first sight. However, Reichel *et al.* performed their experiments in a UHV deposition system, whereas our layers were fabricated in a standard HV system. The resulting higher pressure can lead to contamination during the oxidation process, e.g., by the adsorption of $N_2$ or hydrocarbon molecules. Such adsorbates can reduce the thermodynamic driving force for the transition to the crystalline phase and hamper oxygen uptake leading to oxygen deficient layers. These factors can inhibit the formation of $Al_2O_3$ grains that are large enough to be observed by HRTEM. Due to the non-negligible thickness of TEM samples, individual atoms cannot be resolved and the visibility of grains with sizes of only a few nm strongly depends on their orientation and the TEM specimen thickness. However, if bond angles and coordination numbers in an $AlO_x$ layer deviate only slightly from $Al_2O_3$, such layers will show ELNES features of the crystalline state but still look amorphous in HRTEM images.

Overall, we conclude that the $AlO_x$ layers of EBPlas-250|0.3 and EBPlas-250|9.5 have an amorphous structure as they are still oxygen deficient (cf. Figure 5). However, the bonding characteristics of individual atoms deviate only marginally from those of crystalline $Al_2O_3$ and

exhibit only slightly broadened ELNES fingerprints compared to the crystalline structure. Thus, the requirements for the transition of larger regions into crystalline grains are almost fulfilled. We suggest that the amorphous/crystalline transition can be triggered by a slightly higher oxygen content or/and improved vacuum conditions.

**C. Comparison of structural and nanochemical properties of thin and thicker $AlO_x$ layers**

$AlO_x$-tunnel barriers in JJs require maximum thicknesses ≤ 2 nm, which can be fabricated by a single dynamic or static oxidation process. Due to experimental limitations, thicker $AlO_x$ layers were used for composition analysis applying an iterative Al-deposition/oxidation process, which may has modified the $AlO_x$ properties leading to differences between thin and thicker $AlO_x$ layers. To assure that the results concerning structural and nanochemical properties of thick $AlO_x$ layers are also valid for tunnel barriers, possible differences and their influence on $AlO_x$ properties will be discussed in this Section.

The most obvious difference is the existence of nanocrystalline Al inclusions, which are only present in the thick $AlO_x$ layers and may have influenced the measured oxygen content. It was shown in Section III.A that the fraction of Al inclusions decreases strongly with increasing oxygen pressure. It is rather high for dynamically oxidized samples (EBPlas, EBPlas-UV), where EELS spectra clearly show the ELNES fingerprint of pure Al. Thus, the measured oxygen content most likely differs from tunnel barriers, which are expected to contain a higher oxygen concentration. For static oxidation at high oxygen pressure (EBPlas-70|9.5 and EBPlas-250|9.5) the amount of Al inclusions is reduced to less than 3 %. Possible effects by Al inclusions should be marginal and below the error for the determination of the chemical composition, which is estimated to be between 5 and 10 %. The content of Al inclusions for samples fabricated by static oxidation at low pressure (EBPlas-70|0.3 and EBPlas-250|0.3) is around 8 % and the real oxygen content could be slightly higher than the measured values (cf.

Figure 5). However, as mentioned in Section III.B, the effect is expected to be small, because the edge onset of pure Al is not detected in the EELS spectrum of the Al-$L_{2,3}$ ionization edge (cf. Figure 4a).

To further analyze possible effects of Al inclusions on composition analyses of thin and thick $AlO_x$ layers, EELS spectra of an $AlO_x$-tunnel barrier were acquired with a probe-corrected transmission electron microscope at the Ernst Ruska Center for Microscopy and Spectroscopy with Electrons (Forschungszentrum Jülich, Jülich, GER). The small probe size of this instrument (~0.8 nm) allows acquiring spectra which are not affected by the signal from the lower and upper Al layers and allows to determine the oxygen content of the $AlO_x$-tunnel barrier layer. Sample EBPlas-250|0.3|JJ (cf. Figure 1) contains a thin tunnel barrier and was fabricated under almost the same oxidation conditions as EBPlas-250|0.3 with a thick $AlO_x$ layer. Although the nominal oxidation parameters for both samples were identical (250 °C and 0.3 mbar), slight deviations occurred due to experimental limitations (cf. Supplementary Information). The thin $AlO_x$ layer was actually oxidized within the temperature interval between 260 and 290 °C compared to an oxidation temperature interval between 220 and 280 °C for EBPlas-250|0.3.

The oxygen content in the center of the tunnel barrier (x = 1.25±0.10) and the thick layer (x = 1.22±0.13) agree well within the error limit. A major difference between thin and thick $AlO_x$ layers is the ratio between 'interface' and 'bulk' regions. Hence, all EELS spectra were acquired in 'bulk' regions in the center of the $AlO_x$ layer, where the influence of the $Al/AlO_x$ interface is the lowest. HRTEM images of EBPlas-250|0.3|JJ (cf. Figure 1) show that even $AlO_x$ tunnel barriers contain a 'bulk' region and the interface region is only about one Al lattice plane thick (~ 0.25 nm). Numerical simulations [54] also show an extended 'bulk' region and only thin interface regions in a 1.5 nm thin $AlO_x$ layer. Thus, the structural and nanochemical $AlO_x$ properties in the center of an AlO$_x$-tunnel barrier are expected to correspond to the AlO$_x$ properties in a thick layer as confirmed by the measured oxygen concentrations.

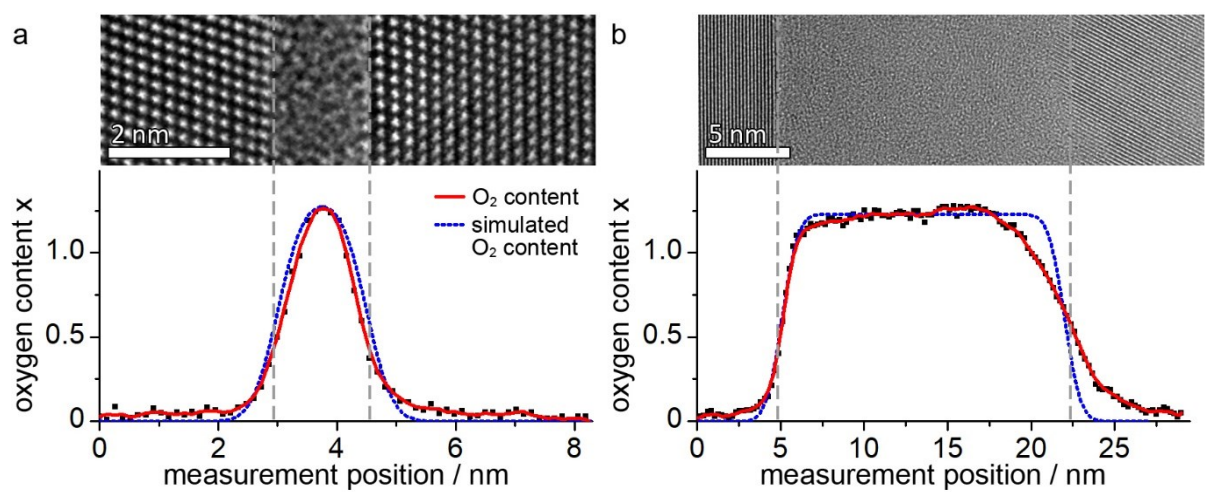

**Figure 6**. HRTEM images and O-concentration profiles obtained from EELS line profiles perpendicular to AlO$_x$ layers fabricated by (a) a single oxidation process (EBPlas-250|0.3|JJ) and (b) iterative oxidation (EBPlas-250|0.3). The O profile in (a) was acquired with a probe-corrected transmission electron microscope in contrast to the O profile in (b) that was taken with a transmission electron microscope without probe-corrector used in all analyses in this work. Blue dotted lines are simulated O-concentration profiles for AlO$_x$ layers with an ideal abrupt Al/AlO$_x$ interface with an electron probe diameter of (a) 0.8 nm and (b) 1.8 nm.

The homogeneity of the oxygen distribution throughout the 'bulk' AlO$_x$ can be visualized by O-concentration profiles obtained from EELS linescans (cf. Figure 6). The oxygen content of the AlO$_x$-tunnel barrier (cf. Figure 6a) shows broadened interface regions of about 1 nm. This is larger than expected from the HRTEM images. However, the diameter of the electron probe of 0.8 nm must be taken into account. The blue curve depicts the simulated O-concentration profile which was obtained by deconvolution of the beam diameter with an O-concentration profile that is characterized by an abrupt chemical Al/AlO$_x$ interface and a homogeneous oxygen content $x = 1.25$. Good agreement between experimental and simulated profiles suggests that the real chemical transition between Al and AlO$_x$ is rather abrupt. The interface region of the thicker AlO$_x$ layer (cf. Figure 6b) is even more diffuse because the profile was acquired in our transmission electron microscope without probe corrector and an electron-beam diameter of ~1.8 nm. Again, the O distribution at the lower Al/AlO$_x$ interface can be well described by the simulated profile (blue line) assuming an abrupt chemical transition at the lower Al/AlOx interface and a probe size of 1.8 nm.

O-concentration gradients are identical at the upper and lower Al/AlO$_x$ interfaces of the AlO$_x$-tunnel barrier layer, in contrast to the broadened O transition at upper AlO$_x$/Al interface of the thicker AlO$_x$ layer. The latter is attributed to the strongly corrugated morphology of the upper interface, which is visible in the HRTEM image Figure 2d. It is assumed to be caused by the iterative Al deposition/oxidation process and does not represent for the real O profile at this interface.

Overall, the oxygen content of thin and thick AlO$_x$ layers grown under the same conditions by static oxidation at high temperature and high oxygen pressure is well comparable with respect to the maximum O concentration in the "bulk" region of the layers. Steep O gradients at Al/AlO$_x$ interfaces occur and a 'bulk' region with constant O distribution is expected even in the thin tunnel barrier. We assume that comparable oxygen contents are obtained in the bulk regions of tunnel barriers and thick layers grown under the same condition in this work. Only crystalline Al inclusions in thick layers may lead to slightly lower O content compared to tunnel barriers. However, even if the absolute O contents may deviate, the effect of temperature and oxygen pressure observed for thicker AlO$_x$ layers – higher temperature and higher oxygen pressure yield a higher oxygen content – will also pertain for the AlO$_x$-tunnel barriers in JJs.

## IV. CONCLUSIONS

The structural and nanochemical properties of AlO$_x$ layers in Al/AlO$_x$/Al-layer systems are correlated with the oxidation parameters in this work. With a thickness of ~2 nm, AlO$_x$ layers can be used as tunnel barriers in superconducting devices, and the results of this work can be considered as guidance to optimize the AlO$_x$ properties in JJs. Our work demonstrates that structure and oxygen content of AlO$_x$ layers can be tuned by the oxidation technique, oxidation temperature and oxygen pressure as summarized in the following:

- Stoichiometric polycrystalline γ-Al$_2$O$_3$ was obtained by electron-beam evaporation of Al$_2$O$_3$ and by reactive sputter deposition. The grain boundaries between crystalline grains contain O$_2$ molecules, which may be a possible source of noise in JJs. The AlO$_x$-layer thickness is mainly controlled by deposition time, which prevents reproducible deposition of layers that are thin enough to be used as tunnel barriers in JJs.

- Amorphous AlO$_x$ was obtained by dynamic and static oxidation. The oxygen content can be tuned by the oxygen pressure and temperature: increasing pressure or/and increasing temperature increases the oxygen content. The highest oxygen content achieved in this work was AlO$_{1.3}$ for $T = 250$ °C and $p = 9.5$ mbar. Dynamic and static oxidation are self-limiting processes, which reproducibly yield AlO$_x$ layers with a thickness of 1.5 – 2.0 nm, making these oxidation techniques ideal for JJs fabrication with high reproducibility.

- Oxidation temperatures above 200 °C lead to significant changes in the bonding characteristics and near-range structural order, which closely resemble crystalline γ-Al$_2$O$_3$, although HRTEM still suggests an amorphous structure. From the observed changes, we expect a significant effect on the resistivity of these layers.

- The same oxygen content was measured in the center of a tunnel barrier and a thick AlO$_x$ layer fabricated with the same oxidation parameters in Al/AlO$_x$/Al-layer systems. This suggests that the correlations between oxidation conditions and AlO$_x$ properties also apply to tunnel barriers in JJs. Absolute oxygen concentrations of thicker AlO$_x$ layers may be slightly lower due to the formation of Al inclusion during the iterative Al deposition/oxidation process.

With respect to reducing noise in JJs, further optimization of the static oxidation process is desirable to induce an amorphous-to-crystalline transition in HV systems, which are typically used for JJ fabrication. Such a transition was already demonstrated by Reichel *et al.* [49] in a UHV system. We suggest that the transition requires a further increase of the O content to approach stoichiometry. Another prerequisite is an epitaxially grown lower Al layer with a (111) surface, where the critical thickness for the amorphous-to-crystalline transition in only in the order of ~0.5 nm. The feasibility of epitaxial growth of Al(111) layer on Si(111) in a HV system was already demonstrated by us [37].

**SUPPLEMENTARY INFORMATION**

**A. Fabrication of Al/AlO$_x$/Al layers**

Three different deposition systems were used to fabricate the Al/AlO$_x$/Al-layer systems, which were analyzed in this work. The majority of the samples were grown in a single-chamber *MEB 550S (PLASSYS Bestek, Marolles-en-Hurepoix, FR)* electron-beam physical vapor deposition system with a base pressure of $10^{-7}$ mbar in the high-vacuum (HV) regime. A sample-plate heater for heating the substrate up to 700 °C and a UV lamp for oxidation enhancement is available in the system. An Ar/O plasma can be generated by a Kaufman source to enhance oxidation or remove carbonaceous contamination.

Before the substrates were mounted in the deposition system, all substrates were consecutively cleaned with N-ethyl-2-pyrrolidon (NEP), isopropyl alcohol and water to remove the protective resist. Samples with Si(001) substrates (EBPlas and EBPlas-UV, cf. Table I of main paper) were then directly transferred into the *Plassys* deposition system where plasma cleaning was applied prior to Al deposition. For the samples fabricated on Si(111) substrates, an additional HF-cleaning process with the buffered oxide etch *BOE 7:1* (12.5 % HF and 87.5 % NH$_4$F) (*Microchemicals GmbH, Ulm, GER*) was performed, before mounting the substrate in the deposition system. Subsequent treatment at temperatures above 700 °C in the deposition system was performed to remove the amorphous SiO$_x$ layer on the substrate and form a reconstructed Si(111) 7x7 surface [60].

Electron-beam evaporation from a pure Al target was used for Al deposition in the *Plassys* at a chamber pressure of $8 - 12 \cdot 10^{-8}$ mbar. The Al layers on Si(001) substrates were deposited without substrate heating at room temperature (~30 °C) and a deposition rate of 0.2 nm/s. Epitaxial Al layers on HF-cleaned Si(111) substrates were obtained using Al-deposition temperatures $\leq 100$ °C and Al-deposition rates $\geq 0.5$ nm/s. These Al layers show significant improvements regarding thickness variations for both Al layer and $AlO_x$ layer with Al-layer thickness variations of less than $\pm 2$ nm and $AlO_x$-layer thickness variations of $\pm 0.11$ nm over lateral distances of ~15 μm.

Different oxidation techniques and parameters were applied to form the $AlO_x$ layer in order to correlate the oxidation conditions with structural and nanochemical properties of the layers. Dynamic oxidation for 12.5 min at room temperature in the *Plassys* was applied for samples EBPlas and EBPlas-UV fabricated on Si(001) substrates with a constant oxygen flow of 10 and 12.7 sccm (cf. Table I, main part of the paper). For EBPlas-UV, additional UV-light illumination was used to enhance the oxidation process. Static oxidation in the *Plassys* was applied for samples deposited on Si(111) with oxygen pressures of 0.3 mbar and 9.5 mbar and substrate temperatures during the oxidation of 70 °C and 250 °C (cf. Table I, main part of the paper). UV-enhanced oxidation in combination with HF-cleaned substrates and static oxidation is not possible in our deposition system, as the UV lamp is positioned in the lid of the load lock and a bake out of the lamp is necessary each time the load lock is opened. This bake out leads to inevitable carbonaceous contamination of the sample, which only can be removed by Ar/O-plasma cleaning with the *Kaufman* source. Unfortunately, plasma cleaning will also oxidize the Si(111) substrate and form a ~3 nm thick amorphous $SiO_x$ layer and thus prevent epitaxial Al growth.

Independent of oxidation conditions, the oxidation process is self-limiting and yields $AlO_x$ layers with a thickness below 2 nm, which is an ideal tunnel-barrier thickness for JJs. However, thicker AlO$_x$ layers are needed for quantitative chemical analysis due to limitations of the used transmission electron microscope (cf. Supplementary Information section B). For this reason, the samples listed in Table I (main part of paper) were produced with a 15 – 20 nm thick AlO$_x$ layer by iterative oxidation. After the deposition of the 100 nm thick lower Al layer and the first oxidation process, 1 nm Al is deposited with a deposition rate of 0.1 nm/s and subsequently oxidized using the same oxidation conditions as for the first oxidation. This process is repeated for up to 15 times until the desired AlO$_x$-layer thickness is reached.

During the oxidation process, considerable temperature variations occurred compared to the pre-set oxidation temperatures given in Table I. During the initial oxidation stage after flooding the chamber with oxygen, the temperature increases for about half a minute and then decreases to the set value. This effect becomes more pronounced with increasing oxygen pressure, which can be explained by enhanced convective heat transfer at higher pressures. At high temperatures, the heat transfer is so strong that the sample plate heater is not capable of keeping the temperature to the set value and the temperature decreases further until the end of the oxidation process. For EBPlas-250|9.5 with a set temperature of 250 °C this results in an actual oxidation temperature range between 300 and 210 °C. Moreover, the oxygen pressure can vary between multiple oxidation steps due to possible time delays and fluctuations of the pressure measurement, which controls the closing of the gas valve. For EBPlas-250|9.5 with a set value of 9.5 mbar the actual pressure varied between 9.3 and 10.0 mbar.

An additional sample, denoted as EBPlas-250|0.3|JJ (nominal conditions: 250 °C and 0.3 mbar, cf. Table I main paper), was fabricated with the same nominal oxidation conditions as EBPlas-250|0.3, but with only a single oxidation step to obtain a ~2 nm thin AlO$_x$-tunnel barrier for comparison between ~20 nm thick AlO$_x$ layers and AlO$_x$-tunnel barriers.

In the last step, a 100 nm thick upper Al layer was deposited using the same deposition parameters as for the lower Al layer.

Two additional samples were grown in other deposition systems for comparison. The sample EBLes (cf. Table I main paper) was grown in a different electron-beam physical vapor deposition system (*PVD 75, Kurt J. Lesker Company, Hastings, UK*) which allows direct $AlO_x$ deposition by evaporation of $Al_2O_3$ pellets. The Al layers were grown with a rate of 0.13 nm/s at a pressure of $1.0 - 1.5 \cdot 10^{-6}$ mbar using a $BN-TiB_2$ crucible with Al pellets. The $AlO_x$ layer with a thickness of ~ 23 nm was also deposited by electron-beam evaporation using a second crucible with $Al_2O_3$ pellets to investigate whether crystalline $Al_2O_3$ can be directly deposited on Al. A deposition rate of 0.03 – 0.04 nm/s at a pressure of $10^{-5}$ mbar and substrate temperature of 75 °C was used without intentional substrate heating or cooling.

Sample ReSput (cf. Table I main paper) was fabricated in a home-built sputter deposition system. It was produced for capacitor fabrication and contains a ~ 25 nm $AlO_x$ layer to achieve a sufficiently high resistivity. The lower and upper Al layers were deposited by Ar-sputtering with an Ar flux of 19 sccm and a rate of 0.5 to 0.6 nm/s at a chamber pressure of $10^{-3}$ mbar. The Al target was also used for sputter deposition of the $AlO_x$ layer with a thickness of about 25 nm by using an Ar/O-plasma (9:1 mixture) at 10 sccm in addition to an increased Ar flux of 33 sccm at a pressure of $1.4 \cdot 10^{-2}$ mbar resulting in a $AlO_x$-deposition rate of 0.45 nm/s.

**B. Transmission electron microscopy and electron energy loss spectroscopy**

Transmission electron microscopy (TEM) was performed with an *FEI Titan³ 80-300* (*Thermo Fisher Scientific, Waltham, USA*) operated at 300 kV. The microscope is equipped with a Gatan imaging filter *Tridiem HR 865* (*Gatan Inc., Pleasanton, USA*) and an aberration corrector in the imaging lens systems. Structural analyses of crystalline regions in the $Al/AlO_x/Al$ layers were performed by comparing two-dimensional (2D) Fourier-transform patterns of high-resolution (HR)TEM images with simulated diffraction patterns using the *JEMS* software [40].

For our experimental setup, which is optimized in respect of a high S/N-ratio for EELS measurements, the diameter of the electron probe (diameter which contains over 90% of the electrons in the beam) in our TEM is ~1.8 nm, which is about the same size than the $AlO_x$-tunnel barrier thickness. Thus, due to unpreventable small sample and beam drifts during the measurement, the electron beam will not only be focused on the $AlO_x$ layer but also interact with the lower and upper Al layer, and EELS spectra will contain signals of both pure Al and $AlO_x$. This effect can be enhanced by an $Al/AlO_x$ interface that is not perfectly parallel to the electron-beam direction, either by misalignment of the sample (more than 0.5° tilt between the electron beam and the $Al/AlO_x$ interface) or by thickness variations of the Al and $AlO_x$ layers. To avoid contributions from the Al layers, EELS spectra were acquired in thicker $AlO_x$ layers, which were fabricated by iterative oxidation and Al deposition. Furthermore, beam-induced damage like carbon contamination, oxygen loss, recrystallization or hole drilling by extended illumination of the same spot with the focused electron beam can be minimized in the thicker $AlO_x$ layers by acquiring spectra in a window of about 3x5 nm² in the center of the $AlO_x$ layer. With the beam continuously scanning over such a region, this leads to a significantly enhanced signal-to-noise ratio in electron energy loss spectra.

Electron energy loss spectroscopy (EELS) was performed in the scanning (S)TEM mode with a convergence angle of 16.7 mrad and a spectrometer acceptance angle of 20.3 mrad. EELS spectra of the $Al-L_{2,3}$ and O-K ionization edges were recorded with acquisition times of 0.01 s up to 10 s and dispersions of 0.02 and 0.05 eV/channel, respectively. A self-written acquisition script collects up to 100 spectra per data point combined with binned-gain averaging to enhance the signal-to-noise ratio. Chemical composition quantification is based on the k-factor method using crystalline γ- and α-$Al_2O_3$ as reference material. Ideally, this method requires the acquisition of the analyzed edges within a single spectrum with high energy resolution (at least 0.1 eV/channel) which is not possible for the $Al-L_{2,3}$ edge (onset at ~76 eV) and the O-K edge (onset at ~ 530 eV) due to the large energy distances between the two ionization edges. Also, the count rate of the O-K edge is lower by two orders of magnitude compared to the $Al-L_{2,3}$ edge. Thus, a script was used that alternatingly acquires 20 spectra of the $Al-L_{2,3}$ and O-K edge with a dispersion of 0.05 eV/channel and varying acquisition times of 20 – 100 ms for the Al-$L_{2,3}$ edge and 2 – 10 s for the O-K edge. The integrated intensity of each edge is measured in energy-loss windows with widths of 30 – 60 eV to minimize effects of the energy-loss near-edge structure (ELNES).

To detect possible differences between the chemical compositions of thin $AlO_x$-tunnel barriers and thicker $AlO_x$ layers (cf. Section III.C main paper), EELS measurements were also performed using a probe-corrected *FEI Titan 80-300* (*Thermo Fisher Scientific, Waltham, USA*) equipped with a Gatan imaging filter *Tridiem 866ERS* (*Gatan Inc., Pleasanton, USA*) at the Ernst Ruska Center for Microscopy and Spectroscopy with Electrons (*Forschungszentrum Jülich, Jülich, GER*). The reduced probe size of 0.8 nm of this instrument allows acquiring EELS spectra in the center of an $AlO_x$-tunnel barrier without any contributions from the lower and upper Al layer. EELS spectra were acquired using the StripeSTEM technique, where a series of 50 to 100 EELS spectra are acquired in a spatially-resolved EELS spectrum image, which is connected to a simultaneously acquired HAADF image. The spectrum image section corresponding to the central part of the $AlO_x$ layer is then used for chemical composition quantification and allows comparison to data obtained in the center of thicker $AlO_x$ layers by the non-probe-corrected transmission electron microscope in our laboratory.

## ACKNOWLEDGMENTS

We thank Dr. Silvia Diewald (CFN Nanostructure Service Laboratory, Karlsruhe Institute of Technology) for carrying out the HF cleaning.

## REFERENCES

[1] M. H. Devoret, R. J. Schoelkopf: "Superconducting circuits for quantum information: An outlook", *Science*, Vol. 339 (6124), 1169–1174, 2013.

[2]  Y.-F. Chen et al.: "Microwave photon counter based on Josephson junctions", *Physical review letters*, Vol. 107 (21), 217401, 2011.

[3]  R. Kleiner et al.: "Superconducting quantum interference devices: State of the art and applications", *Proc. IEEE*, Vol. 92 (10), 1534–1548, 2004.

[4]  P. L. Richards, T.-M. Shen: "Superconductive devices for millimeter wave detection, mixing, and amplification", *IEEE Trans. Electron Devices*, Vol. 27 (10), 1909–1920, 1980.

[5]  T. J. Harvey et al.: "Current noise of a superconducting single-electron transistor coupled to a resonator", *Phys. Rev. B*, Vol. 78 (2), 367, 2008.

[6]  J. Clarke, A. I. Braginski: *The SQUID Handbook - Vol. 1: Fundamentals and technology of SQUIDS and SQUID systems*, Weinheim, Cambridge: Wiley-VCH, 2002.

[7]  M. P. Weides: "Barriers in Josephson junctions: an overview" in *The Oxford Handbook of Small Superconductors*, A. V. Narlikar, Ed.: Oxford University Press, 2017.

[8]  C.-C. Lu et al.: "High mobility flexible graphene field-effect transistors with self-healing gate dielectrics" (eng), *ACS Nano*, Vol. 6 (5), 4469–4474, 2012.

[9]  E. Guerriero et al.: "High-gain graphene transistors with a thin $AlO_x$ top-gate oxide" (eng), *Scientific reports*, Vol. 7 (1), 2419, 2017.

[10] Y. Xuan et al.: "Capacitance-voltage studies on enhancement-mode InGaAs metal-oxide-semiconductor field-effect transistor using atomic-layer-deposited $Al_2O_3$ gate dielectric", *Appl. Phys. Lett,* Vol. 88 (26), 263518, 2006.

[11] D. Shahrjerdi et al.: "Self-aligned inversion-type enhancement-mode GaAs metal-oxide-semiconductor field-effect transistor with $Al_2O_3$ gate dielectric", *Appl. Phys. Lett,* Vol. 92 (20), 203505, 2008.

[12] Y.-T. Wu et al.: "Resistance switching of thin $AlO_x$ and Cu-doped-$AlO_x$ films", *Thin Solid Films*, Vol. 544, 24–27, 2013.

[13] S. Kim, Y.-K. Choi: "Resistive switching of aluminum oxide for flexible memory", *Appl. Phys. Lett,* Vol. 92 (22), 223508, 2008.

[14] D. J. van Harlingen et al.: "Decoherence in Josephson junction qubits due to critical-current fluctuations", *Phys. Rev. B*, Vol. 70 (6), 2004.

[15] J. Burnett et al.: "Noise and loss of superconducting aluminium resonators at single photon energies", *J. Phys.: Conf. Ser,* Vol. 969, 12131, 2018.

[16] S. Sendelbach et al.: "Magnetism in SQUIDs at millikelvin temperatures", *Phys. Rev. Lett,* Vol. 100 (22), 227006, 2008.

[17] S. Choi et al.: "Localization of metal-induced gap states at the metal-insulator interface: origin of flux noise in SQUIDs and superconducting qubits", *Phys. Rev. Lett,* Vol. 103 (19), 197001, 2009.

[18] T. C. DuBois et al.: "Atomic delocalization as a microscopic origin of two-level defects in Josephson junctions", *New J. Phys,* Vol. 17 (2), 23017, 2015.

[19] T. C. DuBois et al.: "Delocalized oxygen as the origin of two-level defects in Josephson junctions", *Phys. Rev. Lett,* Vol. 110 (7), 77002, 2013.

[20] S. Schlör et al.: "Correlating decoherence in transmon qubits: Low frequency noise by single fluctuators"

[21] N. Cabrera, N. F. Mott: "Theory of the oxidation of metals", *Rep. Prog. Phys,* Vol. 12 (1), 163–184, 1949.

[22] A. T. Fromhold, E. L. Cook: "Kinetics of oxide film growth on metal crystals: Electron tunneling and ionic diffusion", *Phys. Rev,* Vol. 158 (3), 600–612, 1967.

[23] C. Ocal et al.: "Cabrera-Mott mechanism for oxidation of metals explains diffusion of metallic atoms through thin defective oxide layers", *Surface Science*, Vol. 163 (2-3), 335–356, 1985.

[24] W.H. Krueger, S. R. Pollack: "The initial oxidation of aluminum thin films at room temperature", *Surface Science*, Vol. 30 (2), 263–279, 1972.

[25] L. P. Jeurgens et al.: "Growth kinetics and mechanisms of aluminum-oxide films formed by thermal oxidation of aluminum", *J. Appl. Phys,* Vol. 92 (3), 1649–1656, 2002.

[26] N. Cai et al.: "Tuning the limiting thickness of a thin oxide layer on Al(111) with oxygen gas pressure", *Phys. Rev. Lett,* Vol. 107 (3), 35502, 2011.

[27] S.'i. Morohashi et al.: "High quality Nb/Al-AlO$_x$/Nb Josephson junction", *Appl. Phys. Lett,* Vol. 46 (12), 1179–1181, 1985.

[28] R. Dolata et al.: "Tunnel barrier growth dynamics of Nb/AlO$_x$ - Al/Nb and Nb/AlN$_x$ - Al/Nb Josephson junctions", *Physica C: Superconductivity*, Vol. 241 (1-2), 25–29, 1995.

[29] N. Cai et al.: "Temperature and pressure dependent Mott potentials and their influence on self-limiting oxide film growth", *Appl. Phys. Lett,* Vol. 101 (17), 171605, 2012.

[30] W. H. Mallison et al.: "Effect of growth conditions on the electrical properties of Nb/Al-oxide/Nb tunnel junctions", *IEEE Trans. Appl. Supercond,* Vol. 5 (2), 2330–2333, 1995.

[31] E. P. Houwman et al.: "Fabrication and properties of Nb/Al, AlO$_x$/Nb Josephson tunnel junctions with a double-oxide barrier", *J. Appl. Phys,* Vol. 67 (4), 1992–1994, 1990.

[32] L. J. Zeng et al.: "Direct observation of the thickness distribution of ultra thin AlO$_x$ barriers in Al/AlO$_x$/Al Josephson junctions", *J. Phys. D: Appl. Phys,* Vol. 48 (39), 395308, 2015.

[33] X. Kang et al.: "Measurements of tunneling barrier thicknesses for Nb/Al–AlO$_x$/Nb tunnel junctions", *Physica C: Superconductivity*, Vol. 503, 29–32, 2014.

[34] L. Zeng et al.: "Atomic structure and oxygen deficiency of the ultrathin aluminium oxide barrier in Al/AlO$_x$/Al Josephson junctions", *Sci. Rep,* Vol. 6, 29679, 2016.

[35] S. Oh et al.: "Elimination of two level fluctuators in superconducting quantum bits by an epitaxial tunnel barrier", *Phys. Rev. B*, Vol. 74 (10), 2006.

[36] M. P. Weides et al.: "Coherence in a transmon qubit with epitaxial tunnel junctions", *Appl. Phys. Lett,* Vol. 99 (26), 262502, 2011.

[37] S. Fritz et al.: "Optimization of Al/AlO$_x$/Al-layer systems for Josephson junctions from a microstructure point of view", *J. Appl. Phys,* Vol. 125 (16), 165301, 2019.

[38] S. Fritz et al.: "Correlating the nanostructure of Al-oxide with deposition conditions and dielectric contributions of two-level systems in perspective of superconducting quantum circuits", *Scientific reports*, Vol. 8 (1), 7956, 2018.

[39] A. Strecker et al.: "specimen preparation for transmission electron microscopy(TEM)- reliable method for cross sections and brittle materials", *Prakt. Metallogr,* Vol. 30 (10), 482–495, 1993.

[40] P. Stadelmann: "Image analysis and simulation software in transmission electron microscopy", *Microsc. Microanal,* Vol. 9 (suppl. 3), 60–61, 2003.

[41] M. Bosman, V. J. Keast: "Optimizing EELS acquisition", *Ultramicroscopy*, Vol. 108 (9), 837–846, 2008.

[42] T. Malis, J. M. Titchmarsh: "'k-factor' approach to EELS analysis" in *Electron Microscopy and Analysis 1985: Inst. Phys. Conf. Ser,* G. J. Tatlock, Ed, Inst. Phys, London-Bristol, 1986, 181–184.

[43] R. F. Egerton: *Electron energy-loss spectroscopy in the electron microscope*, Boston, MA: Springer US, 2011.

[44] M. Heidelmann et al.: "StripeSTEM, a technique for the isochronous acquisition of high angle annular dark-field images and monolayer resolved electron energy loss spectra", *Ultramicroscopy*, Vol. 109 (12), 1447–1452, 2009.

[45] M. J. Verkerk, G. J. van der Kolk: "Effects of oxygen on the growth of vapor-deposited aluminium films", *Journal of Vacuum Science & Technology A: Vacuum, Surfaces, and Films*, Vol. 4 (6), 3101–3105, 1986.

[46] S. Roberts, P. J. Dobson: "The microstructure of aluminium thin films on amorphous SiO$_2$", *Thin Solid Films*, Vol. 135 (1), 137–148, 1986.

[47] R. S. Zhou, R. L. Snyder: "Structures and transformation mechanisms of the η, γ and θ transition aluminas", *Acta Crystallogr B Struct Sci*, Vol. 47 (5), 617–630, 1991.

[48] L.P.H. Jeurgens et al.: "Thermodynamic stability of amorphous oxide films on metals: Application to aluminum oxide films on aluminum substrates", *Phys. Rev. B*, Vol. 62 (7), 4707–4719, 2000.

[49] F. Reichel et al.: "Amorphous versus crystalline state for ultrathin Al$_2$O$_3$ overgrowths on Al substrates", *Journal of Applied Physics*, Vol. 103 (9), 93515, 2008.

[50] R. Brydson: "Multiple scattering theory applied to ELNES of interfaces", *J. Phys. D: Appl. Phys,* Vol. 29 (7), 1699–1708, 1996.

[51] D. Bouchet, C. Colliex: "Experimental study of ELNES at grain boundaries in alumina: Intergranular radiation damage effects on Al-L$_{2,3}$ and O-K edges", *Ultramicroscopy*, Vol. 96 (2), 139–152, 2003.

[52] A. Balzarotti et al.: "Core transitions from the Al 2p Level in amorphous and crystalline Al$_2$O$_3$", *phys. stat. sol. (b)*, Vol. 63 (1), 77–87, 1974.

[53] J. Bruley et al.: "Spectrum-line profile analysis of a magnesium aluminate spinel sapphire interface", *Microsc. Microanal. Microstruct,* Vol. 6 (1), 1–18, 1995.

[54] T. C. DuBois et al.: "Constructing ab initio models of ultra-thin Al/AlO$_x$/Al barriers", *Molecular Simulation*, Vol. 42 (6-7), 542–548, 2015.

[55] P. Glans et al.: "Resonant X-ray emission spectroscopy of molecular oxygen", *Phys. Rev. Lett,* Vol. 76 (14), 2448–2451, 1996.

[56] N. Kosugi et al.: "High-resolution and symmetry-resolved oxygen K-edge spectra of O$_2$", *Chem. Phys. Lett,* Vol. 190 (5), 481–488, 1992.

[57] A. P. Hitchcock, C. E. Brion: "K-shell excitation spectra of CO, N$_2$ and O$_2$", *Journal of Electron Spectroscopy and Related Phenomena*, Vol. 18 (1), 1–21, 1980.

[58] E. Tan et al.: "Oxygen stoichiometry and instability in aluminum oxide tunnel barrier layers", *Phys. Rev. B*, Vol. 71 (16), 2005.

[59] M. Tsuchiya et al.: "Photon-assisted oxidation and oxide thin film synthesis: A review", *Progress in Materials Science*, Vol. 54 (7), 981–1057, 2009.

[60] B. M. McSkimming et al.: "Metamorphic growth of relaxed single crystalline aluminum on silicon (111)", *Journal of Vacuum Science & Technology A: Vacuum, Surfaces, and Films*, Vol. 35 (2), 21401, 2017.